\documentclass[preprint,amsmath,amssymb,nofootinbib]{revtex4}
\usepackage{changes}
\usepackage{amsfonts}
\usepackage{amsmath,graphicx,color,epsfig}

\newcommand{\be}{\begin{equation}}
\newcommand{\ee}{\end{equation}}
\newcommand{\bea}{\begin{eqnarray}}
\newcommand{\eea}{\end{eqnarray}}
\newcommand{\ba}{\begin{array}}
\newcommand{\ea}{\end{array}}
\newcommand{\bi}{\begin{itemize}}
\newcommand{\ei}{\end{itemize}}

\newcommand{\mcb}{{\mathcal B}}

\renewcommand{\vec}[1]{\mbox{\boldmath $#1 \!\!$ \unboldmath}}

\newcommand{\nslash}{\kern 0.2 em n\kern -0.50em /}
\newcommand{\kslash}{\kern 0.2 em k\kern -0.45em /}
\newcommand{\qslash}{\kern 0.2 em q\kern -0.45em /}
\newcommand{\pslash}{\kern 0.2 em p\kern -0.50em /}
\newcommand{\rslash}{\kern 0.2 em r\kern -0.50em /}
\newcommand{\sslash}{\kern 0.2 em s\kern -0.50em /}
\newcommand{\Sslash}{\kern 0.2 em S\kern -0.50em /}
\newcommand{\Pslash}{\kern 0.2 em P\kern -0.50em /}
\newcommand{\Dslash}{\kern 0.2 em D\kern -0.65em /\kern 0.15em}
\newcommand{\lf}{\left}
\newcommand{\rg}{\right}

\begin{document}
\title{The spin parity of $Z_c^-$(4100), $Z_1^+$(4050) and $Z_2^+$(4250)}

\author{Xu Cao}
\affiliation{Institute of Modern Physics, Chinese Academy of Sciences, Lanzhou 730000, China}
\affiliation{
State Key Laboratory of Theoretical Physics, Institute of Theoretical Physics, Chinese Academy of Sciences, Beijing 100190,
China}

\author{Jian-Ping Dai{\footnote{Corresponding author: daijianping@ihep.ac.cn}}}
\affiliation{Experimental Physics Division, Institute of High Energy Physics, Chinese Academy of Sciences, Beijing 100049, China}
\affiliation{INPAC, Shanghai Key Laboratory for Particle Physics and Cosmology, MOE Key Laboratory for Particle Physics,
Astrophysics and Cosmology, Shanghai Jiao-Tong University, Shanghai, 200240, China}

\begin{abstract}
  We conjecture that $Z_c^-$(4100) found by LHCb group from a Dalitz plot analysis of $B^0\to \eta_c K^+\pi^-$ decay is the charge conjugate of $Z_1^+$(4050) observed in $\chi_{c1}\pi^+$ distribution from Belle collaboration. Some interesting conclusions are inferred from this assumption. The $Z_2$(4250) would be assigned to be a $J^P = 1^+$ or $1^-$ state because of its absence in $\eta_{c}\pi^-$ invariant mass distribution, while $Z_1^+$(4050)/$Z_c^-$(4100) could be a $0^+$ or $1^-$ state but $2^+$ is unfavored because it would be coupled to $\eta_{c}\pi$ in $D$-wave. The null observation of $Z_1 Z_2$, $Z_1 Z_1$ and $Z_2 Z_2$ production in $e^+ e^-$ annihilation and $\Upsilon(1S,2S)$ decay by Belle collaboration would further allocate the spin parity combination of $Z_1^+$(4050)/$Z_c^-$(4100) and $Z_2$(4250). Our deductions can be used to exclude a set of proposed models and could be further tested by future experiment, e.g. in $\gamma \gamma$ collisions.

\end{abstract}
\maketitle


The study of new hadronic states has been stimulated by the experimental observation of plenty of exotic structures in the past decade. The unexpected charged states in the heavy
quarkonium sector are especially interesting because it is difficult to incorporate them within the conventional quark-antiquark picture. Several review papers have recently summarized the theoretical and experimental efforts on deciphering their inner components and exploration of relevant underlying dynamics~\cite{Chen:2016qju,Guo:2017jvc,Lebed:2016hpi,Esposito:2016noz,Olsen:2017bmm}. These studies open up a new era of multiquark hadron spectroscopy. The determination of the spin parity of these charmonium-like and bottomonium-like states is in the first place because it is key for our understanding of their nature.

Ten years ago Belle collaboration observed two charged resonance-like structures with the significance of more than 5$\sigma$ in the $\chi_{c1} \pi^+$ mass distribution in $\bar{B}^0 \to K^- \chi_{c1} \pi^+$ decay. Their Breit-Wigner (BW) masses and widths are, respectively~\cite{Mizuk:2008me}
\bea
Z_1^+(4050)&:& \quad 4051 \pm 14^{+20}_{-41} \,\textrm{MeV}, \quad 82^{+21+47}_{-17-22} \,\textrm{MeV}
\\
Z_2^+(4250)&:& \quad 4248^{+44+180}_{-29-35} \,\textrm{MeV}, \quad 177^{+54+316}_{-39-61} \,\textrm{MeV}
\eea
Later BABAR collaboration concluded an absence of signal by analyzing the lower statistical data of $\bar{B}^0 \to K^- \chi_{c1} \pi^+$ and $B^+ \to K_S^0 \chi_{c1} \pi^+$ decays with a detailed study of the acceptance and possible kinematical reflections~\cite{Lees:2011ik}. In the year of 2003, an unpublished thesis from LHCb also claimed non-existence of resonances in ${B}^0 \to K^+ \chi_{c1} \pi^-$ with more than twice the Belle and BABAR cumulative events by using the same analysis strategy with that of BABAR~\cite{Sbordone:2013exb}. While other two experiments made contrary conclusions with Belle, the data themselves in fact agree with each other within uncertainties. The null result in the BABAR and LHCb data would be associated with different treatment of the background.

The mass of $Z_2^+(4250)$ is close to the $D_1 D$ and $D_0 D^*$ threshold. This motivate the interpretation of $D_1 D$ molecular state with QCD sum rules~\cite{Lee:2008gn,Lee:2008tz}. However, the meson exchange model, combining with heavy quark symmetry and chiral symmetry, concludes that the $Z_2^+(4250)$ cannot be a $D_1 D$ or $D_0 D^*$ molecule with reasonable parameters~\cite{Ding:2008gr}. On the other hand, the assignment of $1^-$ tetraquark state is supported by a relativistic diquark-antidiquark picture~\cite{Ebert:2008kb}, QCD sum rule~\cite{Wang:2008af}, and a color flux-tube model with multibody confinement potential~\cite{Deng:2015lca,Deng:2017xlb}. The mass calculated in a QCD sum rule disfavors assigning the $Z_2^+(4250)$ as the compact $0^+$ tetraquark in diquark-antidiquark type~\cite{Wang:2013llv}.

The $Z_1^+$(4050) locates closely to the $D^*\bar{D}^*$ threshold. However, the isotriplet $D^*\bar{D}^*$ molecular interpretation is not favored by QCD sum rule~\cite{Lee:2008gn,Lee:2008tz} and chiral SU(3) quark model~\cite{Liu:2008mi}. The meson exchange model, whose exchanged mesons include pseudoscalar, scalar and vector mesons, find an unbound~\cite{Ding:2009zq} or loosely bound~\cite{Liu:2008tn} $D^*\bar{D}^*$ potential. In latter the $Z_1^+$(4050) is suggested to be a $0^+$ molecular state, supported by a recent calculation of QCD sum rule~\cite{Khemchandani:2013iwa}. Alternatively, it is suggested as a $2^+$ $D^*\bar{D}^*$ molecule in a self-consistent quark-model, where the attraction is from the coupling to $J/\psi \omega$ and $J/\psi \rho$ channels~\cite{FernandezCarames:2009zz,Carames:2010zz}. This picture is compatible with an unitaried coupled-channel model with vector-vector interaction in the framework of hidden gauge formalism, where only one $2^+$ state is dynamically generated for the isotriplet state~\cite{Molina:2009ct} \footnote{This conclusion is inapplicable to the bottomonium sector, where other charged $J^P$ states could be dynamically generated~\cite{Cao:2014vma}.}. It is also proposed as a radially excited $1^+$ state with two light quark $q\bar{q}$ pairs generated dynamically in a coupled-channel Schr\"{o}dinger model with the assumption of the mixing of $^1P_1$ and $^3P_1$ states~\cite{Coito:2016ads}. The $Z_1^+$(4050) as tetraquark candidate in diquark-antidiquark type disfavored by the relativistic diquark-antidiquark picture~\cite{Ebert:2008kb} and a QCD sum rule~\cite{Wang:2013llv}. But it would be a compact $1^+$ tetraquark state in the color flux-tube model with a multibody confinement potential~\cite{Deng:2017xlb}.

Very recently LHCb collaboration found a resonant state with more than three standard deviations in the $\eta_c \pi^-$ invariant mass spectrum of the $B^0 \to K^+ \eta_c \pi^-$ decay~\cite{Aaij:2018bla}. The BW mass and width are,
\be
Z_c^-(4100): \quad 4096 \pm 20^{+18}_{-22} \,\textrm{MeV}, \quad 152 \pm 58^{+60}_{-35} \,\textrm{MeV}
\ee
It is noted that the masses and widths of $Z_c^-$(4100) and $Z_1^+$(4050) are consistent within 1.5$\sigma$ and 1.0$\sigma$, respectively. In this energy range, no other charged states are found at present. It is naturally speculate that they are the same state. No peak with higher mass is observed for $Z_2^+(4250)$. In Table~\ref{tab:JPC}, we list the possible spin-parity $J^{P}$ with low relative orbital quantum number $L$. We do not consider higher partial waves due to their strong suppression. The $C$-parity + is also listed herein if their neural partners exist and $I^G$ is determined to be $1^-$.

\begin{table}
  \begin{center}
 \begin{tabular}{c|c|c}
\hline\hline
    L  & $\chi_{c1} \pi$  & $\eta_c \pi$   \\
\hline
   S   &     $1^{-+}$     &   $0^{++}$     \\
   P   & $(0,1,2)^{++}$   &   $1^{-+}$     \\
   D   & $(1,2,3)^{-+}$   &   $2^{++}$     \\
  $\vdots$   &   $\vdots$ &   $\vdots$     \\
\hline\hline
   \end{tabular}
  \end{center}
  \caption{ Possible $J^{PC}$ assignment of $\chi_{c1} \pi$ and $\eta_c \pi$ system.
  \label{tab:JPC}}
\end{table}

Under this assumption, the $Z_2^+(4250)$ is probably a $J^{P} = 1^{+}$ state, considering that it is present in the $\chi_{c1} \pi$ system but completely missed in $\eta_c \pi$ spectrum. In this case, it is a similar state to $a_1$ in light quark sector. The $a_1$(1640) is really seen to decay into $f_1(1285) \pi$~\cite{Tanabashi:2018oca}, which has the same quantum number with $\chi_{c1} \pi$ system. This decay channel of other $a_1$ states, i.e. $a_1$(1260) and $a_1$(1420), are strongly limited for the sake of small phase space. However, the $1^-$ assignment favored by various tetraquark models~\cite{Ebert:2008kb,Wang:2008af,Deng:2015lca,Deng:2017xlb} is not completely excluded, because it would be suppressed in $\eta_c \pi$ channel for the $P$-wave coupling while it couples to $\chi_{c1} \pi$ in relative $S$-wave. The unfavorable fact of this assignment is that it decays to $\eta_c \pi$ with larger phase space than $\chi_{c1} \pi$, making up the deficiency of higher $L$. This would enhance the $\eta_c \pi$ partial decay width, unless its coupling strength to $\eta_c \pi$ is small by nature, which needs, however, an \textit{ad hoc} reason. Similar argument could be applied to $2^{+}$ assignment, which couples to $\eta_c\pi$ in $D-$wave, higher than that to $\chi_{c1}\pi$. Its presence in $\chi_{c1}\pi$ system does not favor this assignment to some extent because of the $P-$wave coupling, which is also not discussed in any models in
the literatures. So we suspend this possibility in this paper.

An important conclusion can also be driven about the spin-parity of $Z_1^+$(4050)/$Z_c^-$(4100). First, the assignment of $1^{+}$ and $2^{-}$ are impossible because they are forbidden in $\eta_c \pi$ system. So the $1^+$ state with two $q\bar{q}$ pairs~\cite{Coito:2016ads} and compact $1^+$ tetraquark state in the color flux-tube model~\cite{Deng:2017xlb} are both not supported by our assumption. Second, the $2^{+}$ (or $3^{-}$) is also not a good choice for this state because it is expected to be suppressed in $\eta_c \pi$ system for the relative higher $D$-wave (or $F$-wave). As a result, the $2^+$ $D^*\bar{D}^*$ molecule in the quark-model~\cite{FernandezCarames:2009zz,Carames:2010zz} is not preferred in this prescription. The remaining possible $J^{P}$ of $Z_1^+$(4050)/$Z_c^-$(4100) are $0^{+}$ and $1^{-}$.

We can further try to allocate its $J^{P}$ by the measured branching fractions. The following values are extracted by experiments~\cite{Aaij:2018bla,Lees:2011ik,Mizuk:2008me},
\begin{widetext}
\bea
\mcb (\bar{B}^0 \to K^- Z_1^+(4050)) \times \mcb (Z_1^+(4050) \to \chi_{c1} \pi^+) &=& 3.0^{+1.5+3.7}_{-0.8-1.6} \times 10^{-5} \label{eq:Belle}\\
  &<& 1.8 \times 10^{-5} \qquad \textrm{at 90\% C.L.} \label{eq:BABAR}\\
\mcb (B^0 \to K^+ Z_c^-(4100)) \times \mcb (Z_c^-(4100) \to \eta_c \pi^-) &=& 1.89 \pm 0.64 \pm 0.04^{+0.69}_{-0.63} \pm 0.22 \times 10^{-5} \qquad \label{eq:LHCb}
\eea
\end{widetext}
If charge conjugate relation is considered, we can calculate the ratio to be,
\be \label{eq:Zcbr}
\frac{\mcb (Z_c(4100) \to \eta_c \pi)}{\mcb (Z_1(4050) \to \chi_{c1} \pi)} = 0.63^{+0.50}_{-0.89} \quad < \quad 1.05 ^{+0.54}_{-0.51}
\ee
where the upper limit is deduced from Eq.~(\ref{eq:BABAR}) and Eq.~(\ref{eq:LHCb}). This is roughly consistent with the estimation from the heavy quark spin symmetry (HQSS)~\cite{Zhao:2018xrd,Du:2016qcr,Baru:2019xnh},
\bea \label{eq:HQSS}
    \frac{\mcb (Z_c(4100) \to \eta_c \pi)}{\mcb (Z_1(4050) \to \chi_{c1} \pi)} &\simeq& \frac{1}{3}
      \qquad \mbox{for both $1^-$/$0^+$ $Z_c$}
\eea
A factor of 3 is understandable because the $\eta_c$ and $\chi_{c1}$ are the $S$-wave spin singlet and $P$-wave triplet, respectively. Based on the heavy quark spin symmetry (HQSS), the heavy quarkonia, which differ from each other only in the total spin, could be grouped into the same spin multiplet. As a result, it seems that $1^-$ and $0^+$ for $Z_c(4100)$/$Z_1(4050$ are both consistent with data. However, note that the differences of the phase spaces are yet not considered. Here for the case of $Z_c(4100)$/$Z_1(4050)$ with $1^-$, its coupling to $\eta_c\pi$ and $\chi_{c1}\pi$ decays would be in relative $P$- and $S$-wave as indicated in Table~\ref{tab:JPC}, respectively. Therefore after considering the phase space factor $p^{2L}$, above ratio of branching factions will affected by the factor ${p_{\eta_c}^3}/{p_{\chi_{c1}}}$ ($\sim1.6$~(GeV/c)$^2$), where $p_{\eta_c}$ and $p_{\chi_{c1}}$ denote the three-vector momenta of the final mesons in each channel in the decay rest frame. This phase space difference is small and the ratio in Eq.~(\ref{eq:HQSS}) is expected to change a little. Whereas if $Z_c(4100)$/$Z_1(4050)$ is a $0^+$ state, its decays to $\eta_c\pi$ and $\chi_{c1}\pi$ would be in $S$- and $P$-wave, respectively. Thus the above ratio of branching factions is strongly influenced by a big difference of phase space ${p_{\eta_c}}/{p_{\chi_{c1}}^3}$ ($\sim6.4$~(GeV/c)$^{-2}$). Therefore the ratio of this branching factions for $1^-$ is anticipated to be a bit favored by the present experimental constraint in Eq.~(\ref{eq:Zcbr}). This is compatible with a naive argument that $Z_c(4100)$/$Z_1(4050)$ with $1^-$ decays to $\eta_c \pi$ channel with larger phase space volume, compensating its coupling to $\eta_c \pi$ in relative higher $P$-wave. In this sense, the $Z_1^+$(4050)/$Z_c^-$(4100) state is resemble with the light quark $\pi_1$ states. The $\pi_1$(1440) in fact decays strongly to $\eta \pi$~\cite{Tanabashi:2018oca}, but its decay to $f_1(1285) \pi$ is strongly suppressed for the small phase space. However, since the measured branching fractions have large uncertainties, this assignment shall be taken with caution. The $0^{+}$ is still a good candidate for $J^{P}$ of $Z_1^+$(4050)/$Z_c^-$(4100), analog to $a_0$ states in light quark sector. The decay branching ratios of $a_0$(980) and $a_0$(1450) to $\eta \pi$ are big~\cite{Tanabashi:2018oca}, and their decay to $f_1(1285) \pi$ are not seen due to phase space limitation. The $Z_1^+$(4050)/$Z_c^-$(4100) with $0^{+}$ is favored by the molecule explanation of its nature~\cite{Liu:2008tn,Khemchandani:2013iwa}.


We can estimate the ratio of couplings to $\chi_{c1} \pi$ and $\eta_c \pi$ by Eq.~(\ref{eq:Zcbr}), which avoid the effect of phase space difference compared to the widths. The Lagrangians in the hadronic level within the covariant $L-S$ scheme read as~\cite{Zou:2002ar,Cao:2010km},
\bea
{\cal L}^{1^-}_{\chi_{c1} \pi} &=& g_{\chi_{c1} \pi}  \, M_{Z_c} \, \chi_{c1}^\mu \, \vec{\pi} \cdot \vec{Z}_{c\mu} \\
{\cal L}^{1^-}_{\eta_c \pi} &=& g_{\eta_c \pi} \, \lf( \partial^{\mu}\vec{\pi}\, \eta_c - \vec{\pi} \, \partial^{\mu} \eta_c \rg)\cdot \vec{Z}_{c\mu} \\
{\cal L}^{0^+}_{\chi_{c1} \pi} &=& g_{\chi_{c1} \pi} \, \chi_{c1}^\mu \,  \lf( \partial_{\mu} \vec{\pi} \cdot \vec{Z}_{c} + \frac{p^2_{\chi_{c1}}-p^2_{\pi}}{p^2_{Z_c}} \vec{\pi} \cdot \partial_{\mu} \vec{Z}_{c}\rg) \\
{\cal L}^{0^+}_{\eta_c \pi} &=& \frac{g_{\eta_c \pi}}{M_{Z_c}}\, \partial^{\mu} \eta_c \, \partial_{\mu} \vec{\pi} \cdot \vec{Z}_{c} 
\eea
with the dimensionless coupling constants $g_{\chi_{c1} \pi}$ and $g_{\eta_c \pi}$. Then the decay widths can be calculated as,
\bea
{\Gamma}^{1^-}_{\chi_{c1} \pi} &=& \frac{1}{3} \frac{g_{\chi_{c1} \pi}^2}{8\,\pi}\, p_{\chi_{c1}} \, (3+\frac{p_{\chi_{c1}}^2}{M_{\chi_{c1}}^2})\\
{\Gamma}^{1^-}_{\eta_c \pi} &=& \frac{1}{3} \frac{g_{\eta_c \pi}^2}{2\,\pi}\, \frac{p_{\eta_c}^3}{M_{Z_c}^2} \\
{\Gamma}^{0^+}_{\chi_{c1} \pi} &=& \frac{g_{\chi_{c1} \pi}^2}{8\,\pi} \, p_{\chi_{c1}}^3 \frac{\lf(M_{Z_c}^2 + M_{\chi_{c1}}^2 - M_{\pi}^2 \rg)^2}{M_{Z_c}^4\,M_{\chi_{c1}}^2}  \\
{\Gamma}^{0^+}_{\eta_c \pi} &=& \frac{g_{\eta_c \pi}^2}{8\,\pi}\,p_{\eta_c} \frac{(E_{\eta_c} E_{\pi} + p_{\eta_c}^2)^2}{M_{Z_c}^4} 
\eea
where $E_{\pi}$, $E_{\eta_c}$, and $E_{\chi_{c1}}$ denote the energies of the final mesons in each channel in the decay rest frame. Then the following ratio can be computed by Eq.~(\ref{eq:Zcbr}) for,
\bea
    \lf| \frac{g_{\eta_c \pi}}{g_{\chi_{c1} \pi}} \rg| &=&
   \begin{cases}
    2.20^{+1.75}_{-3.11} \quad < \quad 2.85^{+1.47}_{-1.38} &\mbox{for $1^-$ $Z_c$}\\
    0.66^{+0.52}_{-0.93} \quad < \quad 0.85^{+0.44}_{-0.41} &\mbox{for $0^+$ $Z_c$} 
   \end{cases}
\eea
where the uncertainty of $Z_1^+$(4050)/$Z_c^-$(4100) mass is not considered. As can be seen above, the central values of couplings of $Z_1^+$(4050)/$Z_c^-$(4100) to $\chi_{c1} \pi$ and $\eta_c \pi$ are roughly in the same magnitude, however, they are not well confined due to the large uncertainties. Anyway, above ratio is an important clue for the future exploration.

It is worth pointing out whether the $Z_1^+$(4050)/$Z_c^-$(4100) is $0^{+}$ or $1^{-}$ can be disentangled by photon-photon collisions because $1^{-}$ is forbidden in these reactions. If it is produced in photon-photon fusions, it is a $0^{+}$ state and their production ratio of $\gamma \gamma \to \eta_c \pi$ to $\gamma \gamma \to\chi_{c1} \pi$ is expected to be the value in Eq.~(\ref{eq:Zcbr}). Its nature would be also pined down by its two-photon decay width. In the beginning of 1980s, it was predicted that, if the $a_0$(980) and $f_0$(980) mesons are taken as four-quark states, their production rates should be suppressed in photon-photon collisions by a factor ten in comparison with them as conventional two-quark $P$-wave states~\cite{Achasov:1981kh}. The measured values do favor their four-quark structure~\cite{Achasov:2009ee}. The two-photon decay widths of ordinary $0^{+}$ states with both light and heavy $q\bar{q}$ are in the range of several keV~\cite{Tanabashi:2018oca}, and that for the $Z_1^+$(4050)/$Z_c^-$(4100) is expected to be in the order of 0.1 keV if it is of exotic nature. We suggest to probe it in $\gamma \gamma$ interactions at $e^+ e^-$ collider at Belle-II or hadron collider at CERN Large Hadron Collider (LHC). The production rates of $X$(4350) in $\gamma \gamma$ interactions at the LHC are just recently studied~\cite{Goncalves:2018hiw}.

Recently, Belle collaboration observed no significant signals in any of the modes of $e^+ e^-$ and $\Upsilon(1S,2S) \to$ $ Z_1^+(4050) Z_1^-(4050)$, $Z_2^+(4250) Z_2^-(4250)$ and $Z_1^+(4050) Z_2^-(4250)+c.c.$~\cite{Jia:2018yjv}. From our above speculation, the $Z_1(4050)$ would be a $1^-$ or $0^+$ state, and $Z_2(4250)$ might be a $1^+$ or $1^-$. In any of the $1^-$ combinations of $Z_1^+(4050) Z_1^-(4050)$ and $Z_2^+(4250) Z_2^-(4250)$ modes, the relative orbital angular momentum are in $P$-wave at least, resulting into less population in $e^+ e^-$ annihilation and $\Upsilon(1S,2S)$ decays. The $Z_1(4050) Z_2(4250)$ is in $S$-wave for $0^+ 1^-$ and $1^- 1^+$ or $P$-wave for $0^+ 1^+$ and $1^- 1^-$ assignment. Then the latter combinations would be favored by Belle's null results.

In summary, we speculate that $Z_c^-$(4100) found by LHCb group is the charge conjugate state of $Z_1^+$(4050) from Belle collaboration. This possibility is also mentioned in a newly released paper when it discusses the correlations of $Z_c^-$(4100) with some existing exotic candidates~\cite{Zhao:2018xrd}. They are locating far higher than the $D\bar{D}$ open charm threshold, and it is very difficult to incorporate them in the molecular or meson-meson scenario. Their nature is still under wide discussion, including the hadro-charmonium~\cite{Voloshin:2018vym}, tetraquark state~\cite{Wang:2018ntv,Wu:2018xdi} and final state interaction effects~\cite{Zhao:2018xrd} \textit{et al}. We collect all the up-to-date experimental information about them and suggest that spin-parity of $Z_2$(4250) would be $1^+$ or $1^-$, and $Z_1^+$(4050)/$Z_c^-$(4100) is probably $1^-$ or $0^+$, respectively. The first one is more preferred by the measured branching ratios. The null results of double charged charmonium-like state production in $e^+ e^-$ annihilation and $\Upsilon(1S,2S)$ decays from Belle favors that $Z_1^+$(4050)/$Z_c^-$(4100) and $Z_2$(4250) are probably $0^+$ and $1^+$, respectively, or both $1^-$. Though other attributions are not completely excluded due to the limited information, our conjecture about the spin and parity can be used to differentiate various models. Our assignment, together with the extracted ratio of couplings, can be tested by future experiment and give the hint to their nature. We point out that their production in photon-photon collisions would be an ideal place for finally pinning down their spin parity. From another aspect, if one of our conclusion is denied by future experiment, then the $Z_c^-$(4100) and $Z_1^+(4050)$ are not the same.


\begin{acknowledgments}

One of authors (J.-P. Dai) would like to thank Professor Haibo Li for the hospitality during the stay at IHEP, where part of this work has been done. This work was supported by the National Natural Science Foundation of China (Grants Nos. 11405222 and 11505111) and the Key Research Program of Chinese Academy of Sciences (Grant NO. XDPB09).

\end{acknowledgments}

\end{document}